\documentclass[10pt,twoside]{article}

\usepackage[dvips]{graphics,color}
\usepackage{Latex-document}
\usepackage{amssymb}
\newcommand{\bc}{\begin{center}}
\newcommand{\ec}{\end{center}}
\newcommand{\ba}{\begin{eqnarray*}}
\newcommand{\ea}{\end{eqnarray*}}
\newcommand{\bae}{\begin{eqnarray}}
\newcommand{\eae}{\end{eqnarray}}
\newcommand{\bq}{\begin{equation}}
\newcommand{\eq}{\end{equation}}
\newcommand{\ra}{\rightarrow}
\newcommand{\iy}{\infty}
\newcommand{\pr}{\textrm{P}}
\newcommand{\ov}{\over}
\newcommand{\ai}{\textrm{Ai}}

\newcommand{\la}{\lambda}

\newcommand{\be}{\beta}

\def\volumeno{I}
\begin{document}
\markboth{Distribution Functions for Largest Eigenvalues and Their
Applications}{Craig A. Tracy \quad Harold Widom}
\title{\bf  Distribution Functions for  Largest \vskip -2mm Eigenvalues  and
Their Applications\vskip 6mm}
\author{{\bf Craig A. Tracy}\thanks{Department of Mathematics,
University of California, Davis, CA 95616, USA. E-mail:
tracy@math.ucdavis.edu} \quad Harold Widom\thanks{Department of
Mathematics, University of California, Santa Cruz, CA 95064, USA.
E-mail: widom@math.ucsc.edu}\vspace*{-0.5cm}}
\date{\vspace{-8mm}}

\maketitle

\thispagestyle{first} \setcounter{page}{587}

\begin{abstract}\vskip 3mm
It is now believed that the limiting distribution function of the largest
eigenvalue in the three classic random matrix models GOE, GUE and GSE
describe new universal limit laws for a wide variety of processes arising
in mathematical physics and interacting particle systems.  These distribution
functions, expressed in terms of a certain Painlev\'e II function, are described
and their occurences  surveyed.

\vskip 4.5mm


\end{abstract}

\vskip 12mm

\section{Random matrix models} \label{section 1}\setzero

\vskip-5mm \hspace{5mm}

A random matrix model is a
probability space $\left(\Omega,\mathcal{P},\mathcal{F}\right)$
where
the sample space $\Omega$ is a set of matrices.  There are three classic finite $N$ random
matrix models (see, e.g.~\cite{mehta} and for early history~\cite{porter}):
\begin{itemize}
\item Gaussian Orthogonal Ensemble ($\beta=1$)
\begin{itemize}
\item $\Omega=N\times N$ real symmetric matrices
\item $\mathcal{P}$ = ``unique'' measure that is  invariant
under orthogonal  transformations and the matrix elements are i.i.d.\  random
variables.  Explicitly, the density is
\bq c_N\exp\left(-\textrm{tr}(A^2)\right) dA, \label{measure}\eq
where $c_N$ is a normalization constant and $dA=\prod_{i} dA_{ii}\prod_{i<j} dA_{ij}$,
the product Lebesgue measure on the independent matrix elements.
\end{itemize}

\item Gaussian Unitary Ensemble ($\beta=2$)
\begin{itemize}
\item $\Omega= N\times N$ hermitian matrices
\item $\mathcal{P}$= ``unique'' measure that is invariant
under unitary transformations and the (independent) real and imaginary matrix
elements are i.i.d.\  random variables.
\end{itemize}

\item Gaussian Symplectic Ensemble ($\beta=4$) (see \cite{mehta} for a definition)
\end{itemize}

Generally speaking, the interest lies in the $N\ra\iy$ limit of these models.  Here
we concentrate on one aspect of this limit.
In all three
models the eigenvalues, which are
random variables, are real and with probability one they are distinct. If $\la_{\textrm{max}}(A)$
denotes the largest eigenvalue of
the random matrix $A$,  then for each of the three Gaussian
ensembles we introduce the corresponding
distribution function
\[ F_{N,\beta}(t):=\pr_\be\left(\la_{\textrm{max}}< t\right), \beta=1,2,4.\]

The basic limit laws~\cite{tw0, tw1, tw2} state that\footnote{Here
$\sigma$ is the standard deviation of the Gaussian distribution on the
off-diagonal matrix elements.
For  the normalization we've chosen,  $\sigma=1/\sqrt{2}$; however,
for subsequent comparisons, the normalization $\sigma=\sqrt{N}$ is
perhaps more natural.}
\[ F_\be(s):=\lim_{N\ra\iy}F_{N,\be}\left(2\sigma \sqrt{N} +{\sigma s\ov N^{1/6}}\right),\, \be=1,2,4,\]
exist and are given explicitly by
\ba
 F_2(s)&=&\det\left(I-K_{\textrm{Airy}}\right)\\
 \noalign{\vspace{2ex}}
 &=&\exp\left(-\int_s^\infty
(x-s) q^2(x)\, dx\right)
\ea
\par
where
\ba K_{\textrm{Airy}}&\doteq &
{\ai(x) \ai'(y)-\ai'(x)\ai(y)\ov x-y}\\
\noalign{\vspace{1ex}}
&&  \textrm{acting on}\; L^2(s,\iy)\: (\textrm{Airy kernel}) \ea
and $q$ is the unique solution to the Painlev\'e II equation
\[ q^{\prime\prime}=s q + 2 q^3 \]
satisfying the condition
\[ q(s) \sim \textrm{Ai}(s)\:\:\textrm{as}\:\: s\ra\iy.\]

The orthogonal and symplectic distribution functions are
\ba F_1(s) &=& \exp\left(-{1\ov 2}\int_s^\iy q(x)\,dx\right)\, \left(F_2(s)\right)^{1/2},\\
F_4(s/\sqrt{2}) &=& \cosh\left({1\ov 2}\int_s^\iy q(x)\,dx\right)\,
\left(F_2(s)\right)^{1/2}\, . \ea Graphs of the densities $dF_\be/ds$
are in the adjacent figure and some statistics of $F_\be$ can be found
in the table.
\bc \resizebox{10cm}{10cm}{\includegraphics{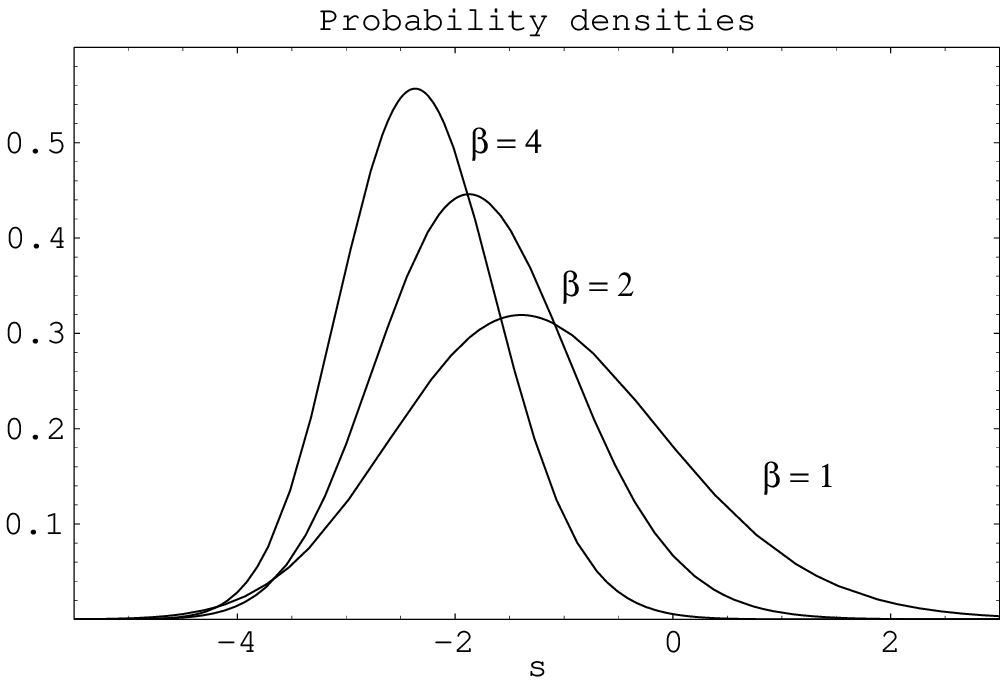}}
\ec
\begin{table}
\begin{center}
\begin{quotation}
\caption{The mean ($\mu_\beta$),  standard deviation ($\sigma_\beta$),
skewness ($S_\beta$) and  kurtosis ($K_\beta$) of $F_\beta$.}
\end{quotation}
\vspace{4ex}
\begin{tabular}{|l|cccc|}\hline
$\beta$ & $\mu_\beta$ & $\sigma_\beta$ & $S_\beta$ & $K_\beta$ \\  \hline
1 & -1.20653 & 1.2680 & 0.293 & 0.165 \\
2 & -1.77109 & 0.9018 & 0.224 & 0.093 \\
4 & -2.30688 & 0.7195 & 0.166 & 0.050 \\ \hline
\end{tabular}
\end{center}
\end{table}

The Airy kernel is an example of an \textit{integrable integral operator}~\cite{IIKS}
and a general theory is developed in \cite{tw3}.
A vertex operator approach to these
distributions (and many other closely related distribution functions in random matrix theory)
was initiated by Adler, Shiota and van Moerbeke~\cite{asm}.  (See the
review article~\cite{vanM} for further developments of this latter approach.)

Historically, the discovery of the connection between Painlev\'e functions
($P_{III}$ in this case)
and Toeplitz/Fredholm determinants appears in work of
 Wu et al.~\cite{wmtb} on the spin-spin
correlation functions of the two-dimensional Ising model.
Painlev\'e functions first appear in random matrix theory in
Jimbo et al.~\cite{jmms} where they prove the Fredholm determinant of the
sine kernel is expressible in terms of $P_V$.  Gaudin~\cite{gaudin}
 (using Mehta's~\cite{mehta0}
then newly invented method of orthogonal polynomials) was the first to discover
the connection between random matrix theory and Fredholm determinants.

\subsection{Universality theorems}

\vskip-5mm \hspace{5mm}

A natural question is to ask
 whether the above limit laws depend upon the underlying Gaussian assumption on
the probability measure.  To
investigate this for unitarily invariant measures ($\be=2$) one replaces
in (\ref{measure})
\[ \exp\left(-\textrm{tr}(A^2)\right)\ra \exp\left(-\textrm{tr}(V(A))\right). \]
Bleher and Its \cite{bleher} choose
\[ V(A)=g A^4 - A^2, g>0, \]
and subsequently   a large class of potentials $V$ was analyzed
by Deift et al.~\cite{dkmvz}.  These analyses require
proving new Plancherel-Rotach type formulas for nonclassical orthogonal
polynomials. The proofs  use Riemann-Hilbert methods.  It was
shown that the generic behavior is   GUE; and hence, the
limit law for the largest eigenvalue
is $F_2$.  However, by finely tuning the potential  new
universality classes will emerge at the edge of the spectrum.
For $\beta=1,4$ a universality theorem was  proved by Stojanovic~\cite{stojanovic}
 for the  quartic potential.

In the case of noninvariant measures, Soshnikov~\cite{soshnikov} proved that
for  real symmetric Wigner matrices\footnote{A symmetric
Wigner matrix is a random matrix
whose entries on and above the main diagonal
are independent and identically distributed
random variables with distribution
function $F$.  Soshnikov
assumes $F$ is even and all moments
are finite.} (complex hermitian Wigner matrices)
the limiting distribution of the largest eigenvalue is $F_1$ (respectively, $F_2$).
The significance of this result is that nongaussian Wigner measures lie outside
the ``integrable class'' (e.g.\ there are no Fredholm determinant representations
for the distribution functions) yet the limit laws are the same as in the integrable
cases.

\section{\boldmath Appearance of $F_\be$ in limit theorems} \label{section 2} \setzero\vskip-5mm \hspace{5mm}

In this section we briefly survey the appearances of the limit laws $F_\be$ in widely differing areas.

\subsection{Combinatorics}

\vskip-5mm \hspace{5mm}

A major breakthrough ocurred with the work of Baik, Deift and Johansson~\cite{bdj} when
they proved that the limiting distribution of the length of the longest increasing subsequence
in a random permutation is $F_2$.  Precisely, if $\ell_N(\sigma)$ is the length
of the longest increasing subsequence in the permutation $\sigma\in S_N$, then
\[ \pr\left({\ell_N - 2\sqrt{N}\ov N^{1/6}} < s\right)\ra F_2(s) \]
as $N\ra\iy$.  Here the probability measure on the permutation group $S_N$ is the uniform measure.
Further discussion of this result can be found in Johansson's contribution to
these proceedings~\cite{joICM}.

Baik and Rains \cite{br1,br2} showed by restricting the set of permutations (and these restrictions
have natural symmetry interpretations) $F_1$ and $F_4$ also appear.  Even the distributions $F_1^2$
and $F_2^2$~\cite{tw4} arise.  By the Robinson-Schensted-Knuth correspondence,
 the Baik-Deift-Johansson result is equivalent
to the limiting distribution on the number of boxes in the first row of random standard Young
tableaux.  (The measure is the push-forward of the uniform measure on $S_N$.)
These same authors conjectured that the limiting distributions of the number of boxes in the second,
third, etc.\ rows were the same as the limiting distributions
 of the next-largest, next-next-largest, etc.\
eigenvalues in GUE.  Since these eigenvalue
distributions were also found in \cite{tw1}, they were able
to compare the then unpublished  numerical work of Odlyzko and Rains~\cite{or}
 with the predicted results of random matrix theory.
Subsequently, Baik, Deift and Johansson \cite{bdj2} proved the conjecture for the second row.
The full conjecture was proved by Okounkov \cite{okounkov} using topological methods and
by Johansson~\cite{jo2} and by Borodin, Okounkov and Olshanski~\cite{boo} using analytical methods.
For an interpretation of the Baik-Deift-Johansson result in terms of the card game
\textit{patience sorting}, see the very readable review paper by Aldous and Diaconis~\cite{ad}.

\subsection{Growth processes}

\vskip-5mm \hspace{5mm}

Growth processes have an extensive history both in the probability
literature and the physics literature (see, e.g.~\cite{gg, meakin, sepp}
and references therein), but it was only recently that Johansson~\cite{jo1, joICM} proved
that the \textit{fluctuations} about the limiting shape in a certain growth model (Corner
Growth Model) are $F_2$. Johansson further pointed out
that certain symmetry constraints (inspired from the Baik-Rains work~\cite{br1,br2}), lead to
$F_1$ fluctuations.  This growth model is in Johansson's contribution to
these proceedings~\cite{joICM} where the close analogy to largest eigenvalue distributions
is explained.

  Subsequently, Baik and Rains \cite{br3} and Gravner, Tracy and Widom \cite{gtw1}
have shown the same distribution functions appearing in closely related lattice growth models.
Pr\"ahofer and Spohn \cite{ps1,ps2} reinterpreted the work of \cite{bdj} in terms of the physicists'
polynuclear growth model (PNG) thereby clarifying the role of the symmetry parameter $\be$.
For example, $\be=2$ describes growth from a single droplet where as $\be=1$ describes
growth from a flat substrate.  They also related the distributions functions $F_\be$ to
fluctuations of the height function in the KPZ equation~\cite{kpz,meakin}.  (The connection
with the KPZ equation is heuristic.)
Thus one expects on physical grounds that the fluctuations
of any growth process falling into the $1+1$ KPZ universality class will be described
by the distribution functions $F_\be$ or one of the generalizations by Baik and Rains~\cite{br3}.
Such a physical conjecture can be tested experimentally; and indeed,
Timonen and his colleagues~\cite{timonen2}
have taken up this challenge.  Earlier Timonen et al.~\cite{timonen1} established
experimentally that a slow, flameless burning process in a random medium (paper!) is
in the $1+1$ KPZ universality class.  This sequence of events
is a rare instance in which new results
in mathematics inspires new experiments in physics.

In the context of the PNG model, Pr\"ahofer and Spohn have given a process
interpretation, the \textit{Airy process}, of $F_2$.  Further work in this direction can
be found in Johansson~\cite{jo4}.

There is an extension of the growth model in \cite{gtw1} to growth in a
\textit{random environment}.
In \cite{gtw2} the following model of interface growth
in two dimensions is considered by introducing a height function on the sites of a one-dimensional
integer lattice with the following update rule: the height above the site $x$ increases to the height
above $x-1$, if the latter height is larger; otherwise the height above $x$ increases by
one with probability $p_x$.  It is assumed that the $p_x$ are chosen independently at
random with a common distribution function $F$, and that the intitial state is such that
the origin is far above the other sites.
In the \textit{pure regime} Gravner-Tracy-Widom identify
an asymptotic shape and prove that the fluctuations about that shape, normalized by
the square root of the time, are asymptotically normal.  This constrasts with the quenched
version: conditioned on the environment and normalized by the cube root of time, the fluctuations
almost surely approach the distribution function $F_2$.
We mention that these same authors in \cite{gtw3}
find, under some conditions on $F$ at the right edge, a \textit{composite regime}
where now the interface fluctuations are governed by the extremal statistics
of $p_x$ in the annealed case while the fluctuations are asymptotically normal in the
quenched case.

\subsection{Random tilings}

\vskip-5mm \hspace{5mm}

The Aztec diamond of order $n$ is a tiling by dominoes of the lattice squares
$[m,m+1]\times[\ell,\ell+1]$, $m,n\in\mathbb{Z}$, that lie inside the region
$\{(x,y): |x|+|y|\le n+1\}$.  A \textit{domino} is a closed $1\times 2$ or $2\times 1$
rectangle in $\mathbb{R}^2$ with corners in $\mathbb{Z}^2$.  A typical tiling is
shown in the accompanying figure.  One observes that near the center the tiling appears random,
called the \textit{temperate zone}, whereas near the edges the tiling is frozen,
called the \textit{polar zones}.  It is a result of Jockush, Propp and Shor \cite{jps}
(see also \cite{cohn}) that as $n\ra\iy$ the boundary between the temperate zone and
the polar zones (appropriately scaled) converges to a circle (Arctic Circle Theorem).
Johansson~\cite{jo3} proved that the fluctuations about this limiting circle are
$F_2$.

\subsection{Statistics}

\vskip-5mm \hspace{5mm}

Johnstone~\cite{joh} considers the largest principal component of the
covariance matrix $X^{t}X$ where $X$ is an $n\times p$ data matrix all
of whose entries are independent standard Gaussian variables and proves
that for appropriate centering and scaling, the limiting distribution
equals $F_1$ in the limit $n,p\ra\iy$ with $n/p\ra\gamma\in
\mathbb{R}^{+}$.  Soshnikov~\cite{sosh2} has removed the Gaussian
assumption but requires that $n-p=\textrm{O}(p^{1/3})$. Thus we can
anticipate applications of the distributions $F_\be$ (and particularly
$F_1$) to the statistical analysis of large data sets.
\begin{table}[th]
\bc \resizebox{10cm}{10cm}{\includegraphics{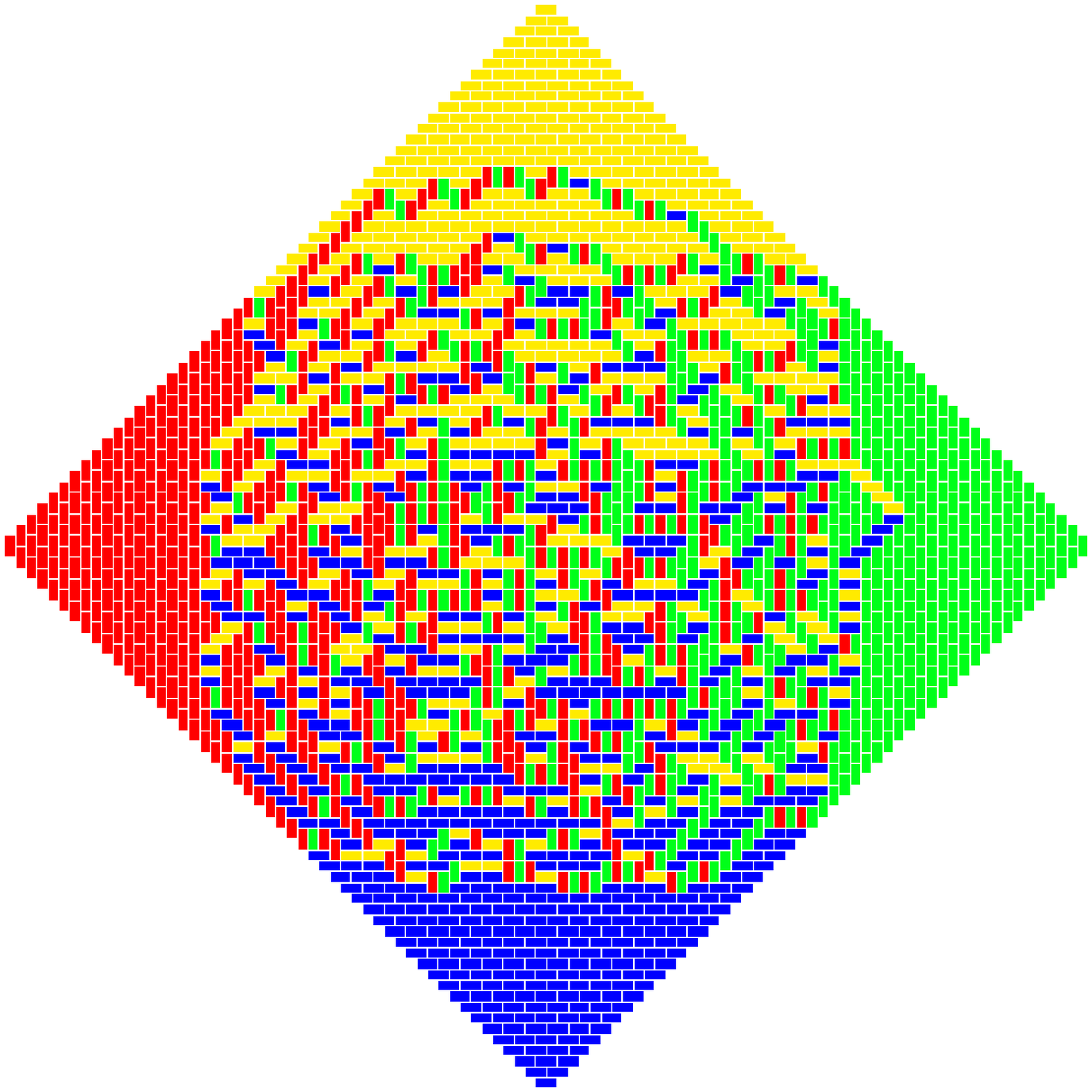}} \vskip 5mm

Random Tilings Research Group \ec
\end{table}

\subsection{Queuing theory}

\vskip-5mm \hspace{5mm}

Glynn and Whitt~\cite{gw} consider a series of $n$ single-server queues each with unlimited waiting
space with a first-in and first-out service.  Service times are i.i.d.\ with mean one
and variance $\sigma^2$ with distribution $V$.  The quantity of interest is $D(k,n)$, the departure time
of customer $k$ (the last customer to be served) from the last queue $n$.  For a fixed number
of customers, $k$,  they prove that
\[ {D(k,n)-n\ov \sigma \sqrt{n}}\]
converges in distribution to a certain functional $\hat D_k$ of $k$-dimensional Brownian motion.
They show that $\hat D_k$ is independent of the service time distribution $V$.  It was
shown in \cite{bar, gtw1} that $\hat D_k$ is equal in distribution to the largest
eigenvalue of a $k\times k$ GUE random matrix. This fascinating connection has been
greatly clarified in recent work of O'Connell and Yor~\cite{oy} (see also \cite{o1}).

From Johansson~\cite{jo1} it follows for $V$ Poisson that
\[ \pr\left({D(\lfloor x n\rfloor,n)-c_1 n\ov c_2\, n^{1/3}}<s\right)\ra F_2(s) \]
as $n\ra\iy$ for some explicitly known constants $c_1$ and $c_2$ (depending upon $x$).

\subsection{Superconductors}

\vskip-5mm \hspace{5mm}

Vavilov et al.~\cite{vav} have  conjectured (based upon certain physical assumptions
supported by numerical work) that the fluctuation of the excitation gap in a metal grain
or quantum dot induced by the proximity to a superconductor is described by $F_1$ for
zero magnetic field and by $F_2$ for nonzero magnetic field.  They conclude their paper
with the remark:
\begin{quote} The universality of our prediction should offer ample opportunities for
experimental observation.
\end{quote}

{\bf Acknowledgements:} This work was supported by the National
Science Foundation through grants DMS-9802122 and DMS-9732687.

\label{lastpage}

\end{document}